\documentclass{elsart5}

\usepackage{graphicx}
\usepackage{amssymb}

\newcommand{\tal}{TlCuCl$_3$}

\begin{document}

\begin{frontmatter}

\title{Thermodynamic properties of the
field-induced N\'{e}el order of TlCuCl\boldmath$_{3}$}

\author[Colonia]{S. Stark}
\author[Colonia]{O. Heyer}
\author[Mos]{A. Vasiliev}
\author[oos]{A. Oosawa}
\author[tan]{H. Tanaka}
\author[Colonia]{T. Lorenz\corauthref{Lorenz}}
\ead{lorenz@ph2.uni-koeln.de}

\address[Colonia]{II.\,Physikalisches Institut, Universit\"{a}t zu
K\"{o}ln,Z\"{u}lpicher Str. 77, 50937 K\"{o}ln, Germany}
\address[Mos]{Department of Low Temperature Physics, Moscow State
University, Moscow 119992, Russia}
\address[oos]{Advanced Science Research Center, Japan Atomic Energy Research
Institute, Tokai, Ibaraki 319-1195, Japan}
\address[tan]{Department of Physics, Tokyo Institute of Technology, Oh-okayama,
Meguro-ku, Tokyo 152-8551, Japan}

\corauth[Lorenz]{ }

\begin{abstract}
\tal\ shows a quantum phase transition from a spin-gap phase to a
N\'{e}el-ordered ground state as a function of magnetic field around
$H_{C0}\simeq 4.8$~T. From measurements of the specific heat $c$
and the linear thermal expansion $\alpha_i$ we calculate the
Gr\"{u}neisen parameter $\Gamma_i(T)=\alpha_i /c$. Close to $H_{C0}$
we find a diverging $\Gamma_i(T\rightarrow 0)$, in qualitative
agreement with theoretical predictions. However, the predicted
individual temperature dependencies of $\alpha(T)$ and $c(T)$ are
not reproduced by our experimental data.

\end{abstract}


\end{frontmatter}


The spin-dimer system \tal\ has been intensively studied in recent
years. Besides the intra-dimer coupling $J\simeq 5.5$~meV there
are also large inter-dimer couplings $J'$
present~\cite{Nikuni2000,matsumoto2002}. The latter cause a
strong dispersion of the triplet excitations, and as a
consequence the minimum singlet-triplet gap $\Delta_m\simeq
0.7$~meV is much smaller than $J$. A moderate magnetic field
$H\simeq 4.8$~T is already sufficient to close $\Delta_m$ and
induces a N\'{e}el order with staggered magnetization perpendicular
to $H$. In the zero-temperature limit, this transition represents
an example of a quantum phase transition~\cite{vojta03}. In the
vicinity of a quantum critical point (QCP) anomalous temperature
dependencies are expected for various physical
properties~\cite{stewart01a}. In particular, a diverging
Gr\"{u}neisen parameter is expected close to a pressure-dependent
QCP~\cite{zhu03a}. Since the phase transition of \tal\ is
extremely sensitive to
pressure~\cite{Oosawa2003a,tanaka2003a,johannsen05a,goto05a} and
the control parameter $H$ may be easily tuned, this compound is
ideally suited to study such generic properties of a QCP.
According to Ref.~\cite{zhu03a} the following temperature
dependencies are expected for $T\rightarrow 0$~K close to the QCP
of \tal :
\begin{equation}
c/T \propto \sqrt{T} \,\, , \,\,\,\, \alpha_i/T \propto 1/\sqrt{T}
\,\, \mbox{ , and } \,\,\,\, \Gamma_i \propto 1/T \,\, .
\label{temps}
\end{equation}
Here, $c$ denotes the specific heat, $\alpha_i$ is the linear
thermal expansion coefficient along the direction $i$ and
$\Gamma_i = \alpha_i /c$ the respective Gr\"{u}neisen parameter.

We present high-resolution measurements of the uniaxial thermal
expansion $\alpha_i=1/L_i \cdot \partial L_i/\partial T$ along
different lattice directions $i$ ($L_i$ is the respective sample
length along $i$) and the specific heat $c$ for $T\gtrsim 0.5$~K.
Since \tal\ easily cleaves along the (010) and (10$\bar{2}$)
planes of the monoclinic structure, we measured $L_i(H,T)$
perpendicular to these planes on a single crystal of dimensions
$1.7 \times 1.5$~mm$^2$ perpendicular to (010) and (10$\bar{2}$),
respectively. In addition, the [201] direction was measured on a
second crystal of length $L_{[201]}=4.4$~mm. All properties have
been studied up to $H=14$~T applied perpendicular to the
(10$\bar{2}$) plane.

\begin{figure}[t]
\begin{center}
\includegraphics[width=.48\textwidth]{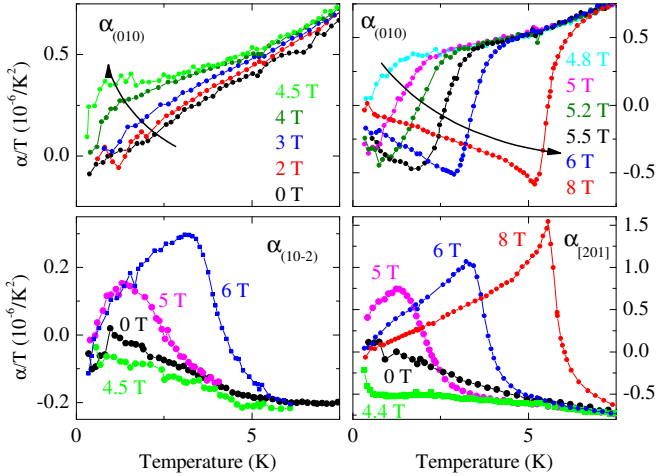}
\end{center}
\caption{Top: Thermal expansion perpendicular to the $(010)$ plane
for different magnetic fields $H<H_{C0}\simeq 4.8$~T (left) and
$H\geq H_{C0}$ (right). The arrows indicate increasing
magnetic-field strength. The bottom panels show some $\alpha_i/T$
curves measured perpendicular to the $(10\overline{2})$ plane
(left) and along the $[201]$ direction (right).} \label{alpT}
\end{figure}

\begin{figure}[b]
\begin{center}
\includegraphics[width=.48\textwidth]{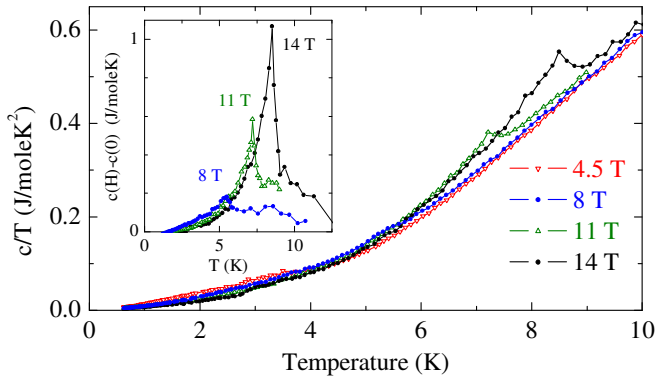}
\end{center}
\caption{Specific heat of \tal\ for different magnetic fields.
The inset shows $c(H)-c(0)$ to visualize the anomalies at
$T_{N}$.}\label{cp}
\end{figure}

Fig.~\ref{alpT} shows $\alpha_i$ measured along all three
directions for different magnetic fields.  In zero field,
$\alpha_i/T$ continuously approaches zero for $T\rightarrow 0$.
With increasing $H$ a pronounced shoulder develops, which reaches
a maximum around 4.5~T. For larger fields clear anomalies with a
sign change of $\alpha_i/T$ occur. These anomalies systematically
sharpen and shift towards higher $T$ with further increasing
field. From magnetostriction measurements (not shown) we find a
critical field $H_{C0}\simeq 4.8$~T for $T\rightarrow 0$. The
broad anomalies appear in the $\alpha_i(T)$ curves already for $H
\gtrsim 4.6$~T. We attribute this to the finite width of the
phase transition. The curves for $i=(010)$ and $(10\bar{2})$
agree well with our previous results measured on a different
crystal for  $T \gtrsim 3$~K~\cite{johannsen05a,johannsen06a}.
There we have also shown that the $\alpha_i$ anomalies signal
huge uniaxial pressure dependencies of $T_N$, which arise from
the pressure-dependent changes of the intra-dimer coupling $J$.
From the sum of all three $\alpha_i$ we can also determine the
volume changes and calculate the hydrostatic pressure dependence
of the spin gap. We obtain $\partial \ln \Delta_m /\partial
p_{hydro} \simeq -360$~\%/GPa, in reasonable agreement with the
initial slope of $\simeq -400$~\%/GPa found by magnetization
measurements under hydrostatic pressure~\cite{goto05a}.

\begin{figure}[t]
\begin{center}
\includegraphics[width=.48\textwidth]{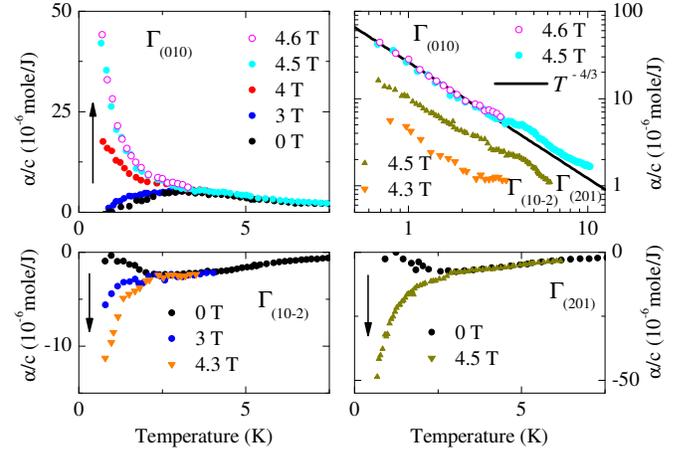}
\end{center}
\caption{Gr\"{u}neisen parameters $\Gamma_i=\alpha_i/c$ for different
fields (the arrows indicate increasing field strength). The upper
right panel shows $\Gamma_i(T)$ for $H\lesssim H_{C0}$ on
double-logarithmic scales. Here, the negative $\Gamma_i(T)$ for
$i=(10\overline{2})$ and $[201]$ have been divided by a factor of
$-2$ and $-3$, respectively. The solid line is a power-law fit
yielding $\Gamma_{(010)}\propto T^{-4/3}$.} \label{Ga}
\end{figure}

In Fig.~\ref{cp} we show the specific heat for different fields.
In agreement with Ref.~\cite{Oosawa2001} we find a rather small
anomaly even for the largest field, and the magnitude rapidly
decreases when $H_{C0}$ is approached. From $\alpha_i$ and $c$ we
calculated the uniaxial Gr\"{u}neisen parameters $\Gamma_i$ shown in
Fig.~\ref{Ga}. For all three directions $\Gamma_i$ remains finite
for $T\rightarrow 0$ in zero and small fields. However, all three
$\Gamma_i$ show a clear tendency to diverge for $H\simeq 4.5$~T.
In the upper right panel of Fig.~\ref{Ga} we display
$\Gamma_i(T)$ for $H\lesssim H_{C0}$ on double-logarithmic
scales. Within experimental accuracy, the slope of all three
$\Gamma_i(T)$ is identical. A power-law fit yields $\Gamma_{(010)}
\propto T^{-4/3}$ and describes the experimental data reasonably
well for about one decade. Since only the irregular contributions
of $c$ and $\alpha_i$ are considered in Ref.~\cite{zhu03a}, while
the experimental data also contain the respective phononic
contributions, one may tend to the conclusion that our data
nicely confirm the theoretical prediction $\Gamma(T)\propto
T^{-1}$. However, the agreement between theory and experiment is
much worse when the individual temperature dependencies of $c/T$
and $\alpha_i/T$ are considered. Neither the data of
Fig.~\ref{alpT} nor those of Fig.~\ref{cp} give any indication to
follow the predicted temperature dependencies of
Eq.~(\ref{temps}). Concerning the specific heat data, one may
argue that the predicted $\sqrt{T}$ behavior is difficult to
identify because of the phononic contribution. However, this
argument does not hold for $\alpha_i$, since the predicted
divergence of $\alpha_i/T$ should be seen despite a (regular)
phononic contribution. Thus, our experimental data do only
partially confirm the theoretical predictions~\cite{zhu03a}.

In summary, we find diverging Gr\"{u}neisen parameters $\Gamma_i
=\alpha_i/c$ close to the quantum critical point of \tal , in
qualitative agreement with theoretical predictions. However,
$\alpha_i$ and $c$ do not show the predicted temperature
dependencies. One may suspect that this disagreement could arise
from the magnetic anisotropy of \tal , which is not considered in
Ref.~\cite{zhu03a}, or from a broadening of the transition due to
sample imperfections, as e.g.\ internal strains.

We acknowledge discussions with A.~Rosch, I.~Fischer, and
J.A.~Mydosh. This work was supported by the DFG via SFB 608.



\end{document}